\documentclass{iaus}
\usepackage{graphicx}
\usepackage{natbib}
\graphicspath{{./fig/}}
\usepackage{bm}

\def\Rm{\mbox{\rm Re}_{\rm M}}
\def\Beq{B_{\rm eq}}
\def\etat{\eta_{\rm t}}

\title[Magnetic helicity fluxes in $\alpha\Omega$ dynamos] %% give here short title %%
{Magnetic helicity fluxes in $\alpha\Omega$ dynamos}

\author[Simon Candelaresi \& Axel Brandenburg]   %% give here short author list %%
{Simon Candelaresi
 \and Axel Brandenburg}

\affiliation{NORDITA, AlbaNova University Center,
Roslagstullsbacken 23, SE-10691 Stockholm, Sweden \\
Department of Astronomy, Stockholm University,
SE 10691 Stockholm, Sweden}

\pubyear{2010}
\volume{274}  %% insert here IAU Symposium No.
\pagerange{464--466}
\jname{Advances in Plasma Astrophysics}
\editors{A. Bonanno, E. de Gouveia dal Pino \& A. Kosovichev, eds.}
\doi{10.1017/S1743921311007502}
\begin{document}

\maketitle

\begin{abstract}
In turbulent dynamos the production of large-scale magnetic fields is
accompanied by a separation of magnetic helicity in scale.
The large- and small-scale parts increase in magnitude.
The small-scale part can eventually work against
the dynamo and quench it, especially at high magnetic Reynolds numbers.
A one-dimensional mean-field model of a dynamo is presented where
diffusive magnetic helicity fluxes within the domain are important.
It turns out that this effect helps to alleviate the quenching.
Here we show that internal magnetic helicity fluxes, even within one
hemisphere, can be important for alleviating catastrophic quenching.

\keywords{Sun: magnetic fields}
\end{abstract}

% \firstsection % if your document starts with a section,
The magnetic fields of astrophysical bodies like stars and galaxies
show strengths which are close to equipartition.
The scale of the magnetic field is larger then the
dissipation length and reaches the order of the size of the object.
The mechanism which creates those fields is believed to be a dynamo.
Large- and small-scale
magnetic helicity with opposite signs are created. For high magnetic
Reynolds numbers, $\Rm$, this makes the dynamo saturate only on a resistive
time scale and reduces the saturation
field strength much below equipartition \citep{BS05c}.
This effect is called catastrophic quenching and increases with increasing Reynolds
number, because
for the Sun $\Rm = 10^{9}$ and for galaxies $\Rm = 10^{14}$.
This suggests that helicity has to be shed.
Observations have shown \citep{Manoharan1996, Canfield1999} that
helical structures
on the Sun's surface are more likely to erupt into coronal mass ejections
(CMEs). This suggests that the Sun sheds magnetic helicity by itself.

\begin{figure}[h]
\begin{minipage}[b]{0.48\linewidth}
\centering
\includegraphics[width=1.0\linewidth]{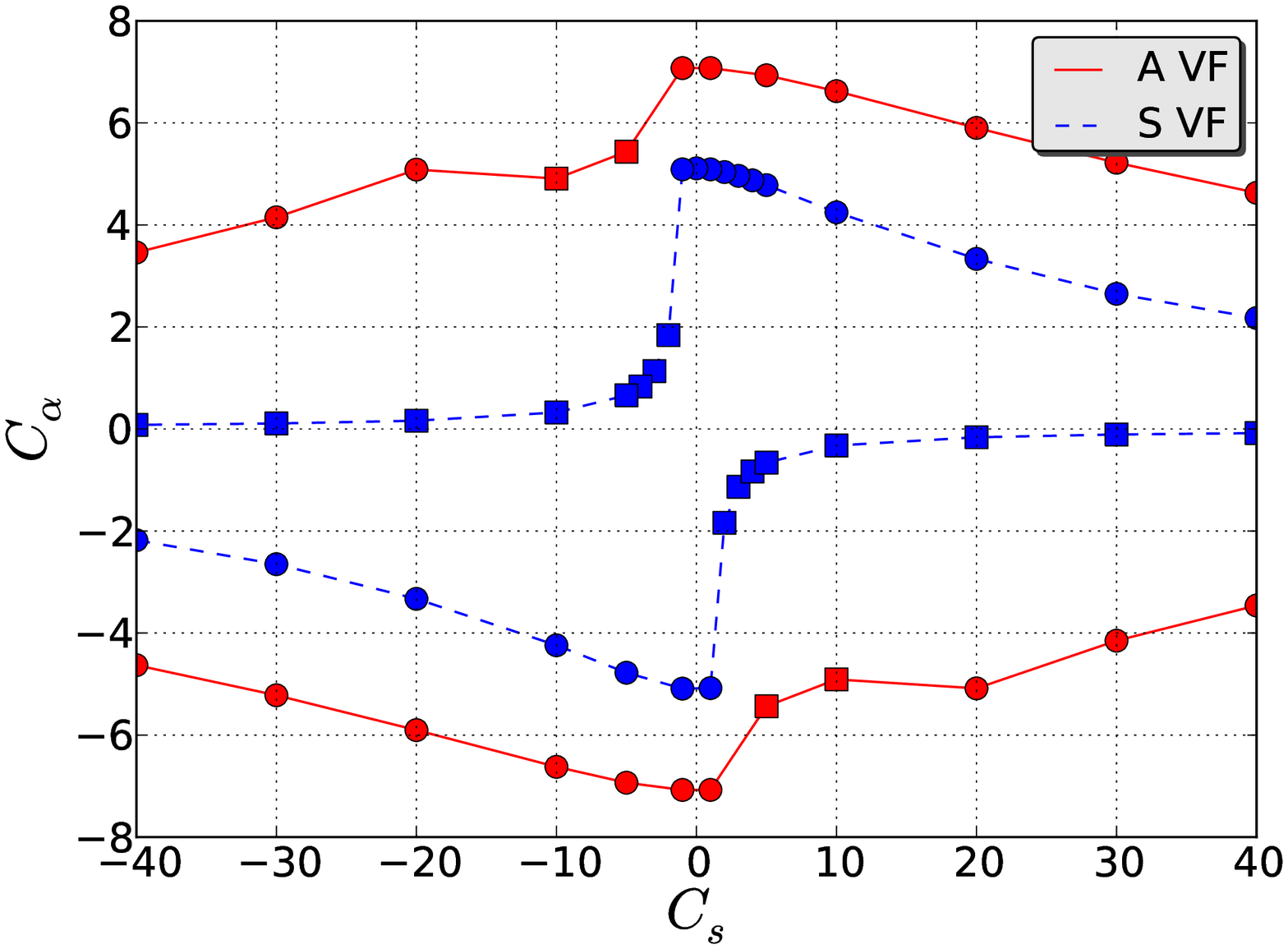}
\caption{Critical values for the strength of the forcing $C_{\alpha}$
and the shear $C_{\rm S}$ for which dynamo action occurs for the cases
of vertical field boundary conditions
and antisymmetric (solid, red) and symmetric (dashed, blue)
equator.
The circles and squares represent oscillating and stationary solutions
respectively.}
\label{fig: lambda crit A}
\end{minipage}
\hspace{0.5cm}
\begin{minipage}[b]{0.48\linewidth}
\includegraphics[width=1.0\linewidth]{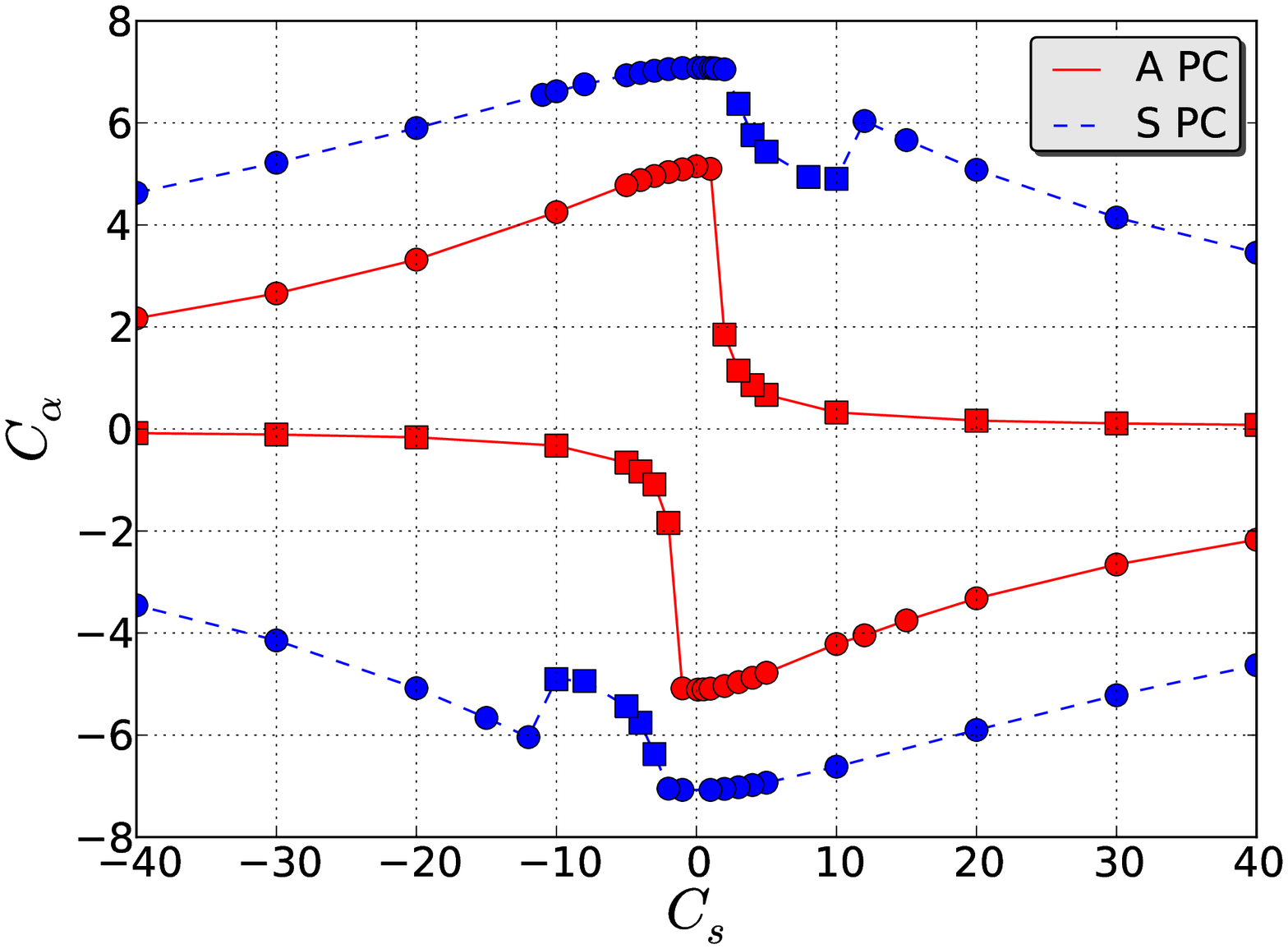}
\caption{Critical values for the strength of the forcing $C_{\alpha}$
and the shear $C_{\rm S}$ for which dynamo action occurs for the cases
of perfect conductor boundary conditions
and antisymmetric (solid, red) and symmetric (dashed, blue)
equator.
The circles and squares represent oscillating and stationary solutions
respectively.}
\label{fig: lambda crit S}
\end{minipage}
\end{figure}

In our earlier work \citep{ssHelLoss09}
we have considered a one-dimensional mean-field model in the $z$-direction of a
dynamo with wind-driven magnetic helicity flux where the wind increases with distance
from the midplane.
Magnetic helicity evolution is taken into account by using what is
known as the ``dynamical quenching'' formalism that is described
in our earlier paper and in references therein.
We augment these studies by imposing a constant
shear throughout the domain which facilitates the growth of the
magnetic energy.
We perform simulations in one hemisphere where we set the magnetic
field in the $z$-direction to be symmetric (S) or antisymmetric (A)
at the midplane.
The outer boundaries are set to either vertical field (VF) or
perfect conductor (PC).
The free parameters are the dynamo numbers $C_{\alpha}$ and
$C_S$. By varying both numbers we find the critical values for
dynamo action (Fig.\,\ref{fig: lambda crit A} and
Fig.\,\ref{fig: lambda crit S}).
The critical values for $C_{\alpha}$ decrease when the shear increases.
This is expected, since larger shear leads to stronger toroidal field
which enhances the dynamo effect.

\begin{figure}[t!]
% SC: reduced size to 0.95 to fit the paper within 3 pages
\centering\includegraphics[width=0.95\columnwidth]{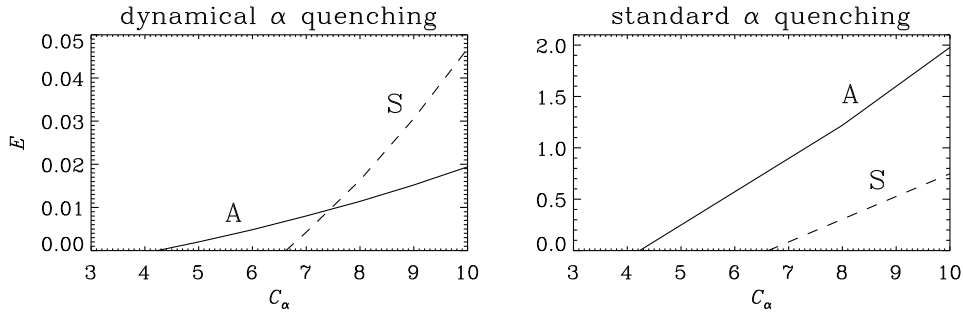}\caption{
Comparison of the bifurcation diagrams for dynamical and standard
$\alpha$ quenching.
}\label{pbif}\end{figure}

\begin{figure}[t!]
\centering\includegraphics[width=\columnwidth]{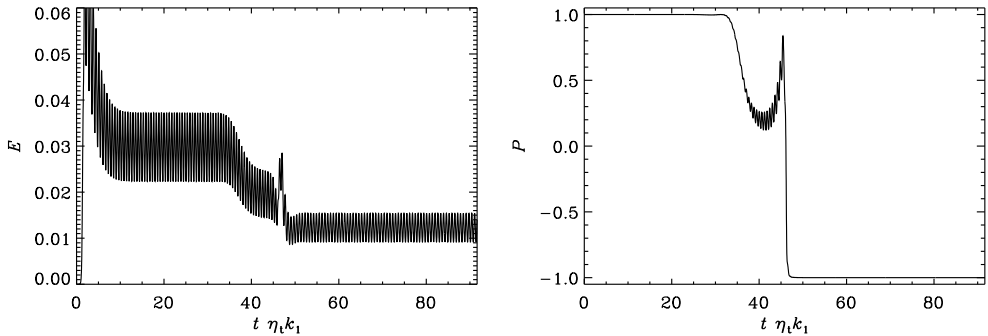}\caption{
Evolution of the magnetic energy (left) and parity (right) for a solution
that was initially even about the midplane (S or quadrupolar solution),
but this solution is unstable and developed an odd parity
(A or dipolar solution).
}\label{pparity_comp}\end{figure}

In the rest of this paper we study in more detail the case $C_S=-10$
and consider positive values of $C_\alpha$.
In Fig.\,\ref{pbif} we compare the dynamical $\alpha$ quenching model (using
a magnetic Reynolds number of $\Rm=10^5$) with
the standard (non-catastrophic) $\alpha$ quenching where $\alpha\propto1/(1+\overline{\bm{B}}^2/\Beq^2)$
with $\overline{\bm{B}}$ being the mean field and $\Beq$ the equipartition value.
Note that in the former case, the energies cross.
Nevertheless, the A solution is stable in both cases -- at least
for $C_\alpha\leq10$.
This is demonstrated in Fig.\,\ref{pparity_comp}, where we show that after
about 40 diffusive times, $\etat k_1 t=40$, where $\etat$ is the
turbulent magnetic diffusivity and $k_1$ the basic wavenumber,
the magnetic energy $E$ decreases and the parity $P$ swaps from $+1$
to $-1$; see \cite{Betal89} for details on similar studies.

\begin{figure}[t!]
\centering\includegraphics[width=\columnwidth]{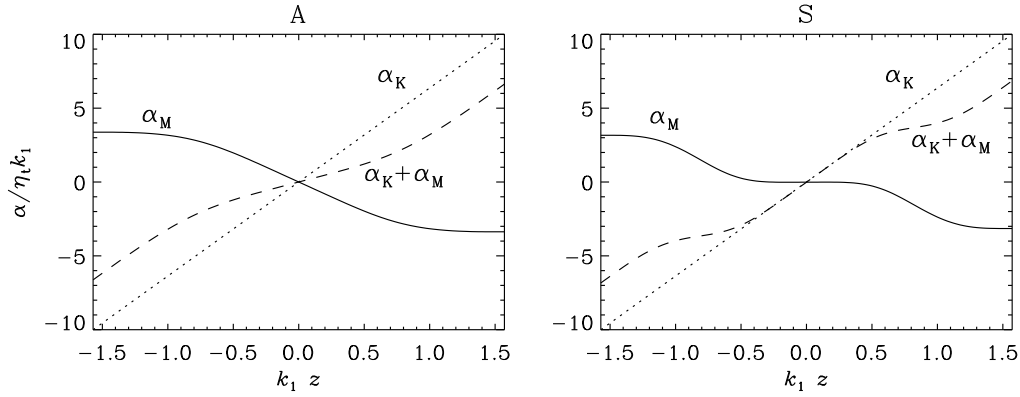}\caption{
Profiles of $\alpha_{\rm K}$, $\alpha_{\rm M}$, and their sum
for the A and S solutions at $C_\alpha=10$.
}\label{palp_comp}\end{figure}

The crossing of the energies in the dynamical quenching model
is somewhat surprising.
In order to understand this behavior, we need to look at the
profiles of the $\alpha$ effect; see Fig.\,\ref{palp_comp}.
In this model, $\alpha$ is composed of a kinetic part, $\alpha_{\rm K}$,
and a magnetic part, $\alpha_{\rm M}$, which has typically the opposite
sign, which leads to a reduction of $\alpha=\alpha_{\rm K}+\alpha_{\rm M}$.
The quenching can be alleviated by reducing $\alpha_{\rm M}$, for example
when the divergence of the magnetic helicity flux of the small-scale
field, $\overline{\bm{F}}_{\rm f}$, becomes important.

\begin{figure}[t!]
\centering\includegraphics[width=\columnwidth]{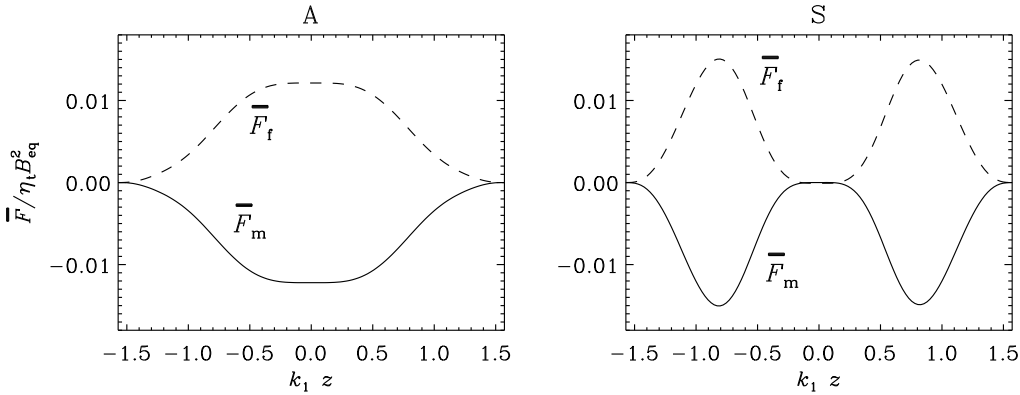}\caption{
Time-averaged magnetic helicity fluxes, $\overline{\bm{F}}_{\rm f}$
and $\overline{\bm{F}}_{\rm m}$, of
fluctuating and mean fields, for the A and S solutions, respectively,
at $C_\alpha=10$.
Note that $\overline{\bm{F}}_{\rm f}+\overline{\bm{F}}_{\rm m}\approx0$.
}\label{pflux_comp}\end{figure}

Naively, we would have expected that the A solution should have a larger
energy, because only this solution allows a magnetic helicity flux through
the equator; see Fig.\,\ref{pflux_comp}.
This is however not the case, which may have several reasons.
Even though the magnetic helicity flux flux small at the equator ($z=0$),
there can be significant contributions from within each hemisphere
which contributes to alleviating the catastrophic quenching.
The details of this will be address in more detail elsewhere.

\newcommand{\yan}[3]{ #1, \textit{AN,} \textrm{#2}, #3}
\newcommand{\yana}[3]{ #1, \textit{A\&A,} \textrm{#2}, #3}

\end{document}